\documentclass[prd,twocolumn,noshowpacs,preprintnumbers,amsmath,amssymb,nofootinbib]{revtex4}
\usepackage{bm,calrsfs,color}
\usepackage[normalem]{ulem} 
\usepackage[colorlinks=true, pdfstartview=FitV, linkcolor=blue, citecolor=blue, urlcolor=blue]{hyperref}

\def\a{\alpha}

\def\m{\mu}
\def\n{\nu}

\def\x{\xi}

\def\be{\begin{equation}}
\def\ee{\end{equation}}
\def\beq{\begin{eqnarray}}
\def\eeq{\end{eqnarray}}

\def\ca{{\cal A}}

\def\ct{{\cal T}}

\def\cw{{\cal W}}

\begin{document}

\title{\LARGE A classification of global conformal invariants}

\author{Nicolas BOULANGER}
\affiliation{\\ Service de Physique de l'Univers, Champs et Gravitation, 
Universit\'e de Mons -- UMONS, 
20 Place du Parc, B-7000 Mons, Belgique}
\author{Jordan FRAN\c{C}OIS}
\affiliation{\\  Universit\'e de Lorraine,CNRS, Institut {\'E}lie Cartan de Lorraine, UMR 7502, 
Vandoeuvre-l\`es-Nancy, F-54506, France}
\author{Serge LAZZARINI}
\affiliation{\\ Aix Marseille Univ, Universit\'e de Toulon, CNRS, CPT, Marseille, France}

\begin{abstract} 

\medskip

\hfill{\em Dedicated to the memory of our colleague Christian Duval 1947--2018.}

\bigskip

\bigskip
We provide the full classification, in arbitrary even and odd dimensions, of global conformal invariants, i.e., scalar densities 
in the spacetime metric and its derivatives that are invariant, 
possibly up to a total derivative, under local Weyl rescalings of the metric. We use cohomological techniques that 
have already proved instrumental in the classification of Weyl anomalies in arbitrary 
dimensions. The approach we follow is purely algebraic and 
borrows techniques originating from perturbative Quantum Field Theory for which locality is crucial.
\end{abstract}

\maketitle


\section{Introduction}

Since Weyl's seminar work of 1918 \cite{Weyl1918}, Weyl invariance in gravitational theories 
has always been a topic of high interest both in Mathematics --- where the terminology in use 
is Conformal Geometry --- and in Theoretical Physics, where recent papers and references can 
be found e.g. in \cite{Scholz:2017pfo}. 
As soon as the conformal symmetry is incorporated in a model, the problem of establishing 
the general structure of conformal invariants immediately arises.
In Physics, the question concerns the classification of  scalar densities 
(Lagrangian densities) built out of the metric and its derivatives that are invariant, 
possibly up to a total derivative, under both diffeomorphisms and local rescalings 
(Weyl transformations) of the spacetime metric.
In Conformal Geometry, this problem amounts to the classification of 
global conformal invariants and has been studied extensively, 
see {\em e.g.} \cite{curry_gover_2018} for some recent references. 

By definition, global conformal invariants are given by the integral
over a $n$-dimensional (pseudo) Riemannian manifold ${\cal M}_{n}(g)$ 
of linear combinations 
of strictly Weyl-invariant scalar densities and scalar densities that are invariant under 
Weyl rescalings only up to a total derivative.
The solution to the above classification problem can be obtained by 
translating it in the language of local field theory  
in the formulation of  Becchi, Rouet, Stora, and Tyutin (BRST) 
\cite{Becchi:1974md,Tyutin:1975qk,DuboisViolette:1985jb}
thereby recognising its purely algebraic nature. Locality means here that we shall deal with 
the space of local functions parametrized by monomials in the derivatives of the fields, 
but not necessarily monomials in the undifferentiated fields (as for example $\sqrt{-g}$, the square 
root of the determinant of the metric), 
along the idea developed by Dixon in the 1970's and published much later in \cite{Dixon:1991wi}, see {\em e.g.} \cite{Bandelloni:1987kg} and references therein.
Therefore, it turns our that the global conformal invariants are indeed given by the cohomology of the BRST differential in form degree $n$ and ghost number zero, in the jet space of the metric and its spacetime derivatives. 
The BRST differential, in its turn, is known from the diffeomorphisms and Weyl 
(or conformal) rescalings of the metric, while  
the ghost number is the $\mathbb{N}$ grading associated with the 
BRST differential. We refer to \cite{Barnich:2000zw,Barnich:1995ap} 
for a modern review and an exhaustive list of references on the BRST formulation in local field theory. 
In this spirit, a similar approach has also been developed earlier for the diffeomorphisms in \cite{Bandelloni:1988ws}.
By the assumption of locality, a  global invariant is given by a 
ghost-zero scalar density whose Hodge dual $a^{0,n}$ in dimension $n$ is a cocycle of the BRST differential $s$ modulo the exterior derivative of a local $(n-1)$-form $b^{1,n-1}$ in ghost number 1. In other words, it must obey the cocycle equation $sa^{0,n} + db^{1,n-1} = 0\,$ as introduced by R.~Stora in \cite{Sto77}.

In this paper, by giving the general structure of the solutions of the equation  
$sa^{0,n} + db^{1,n-1} = 0\,$, we provide a purely cohomological classification of the  global 
conformal invariants in arbitrary dimension $n\,$, be it even ($n=2m$) or odd ($n=2m+1\,$),  
respectively.
As usual in this context, see for example the chapter 9 of 
\cite{Barnich:2000zw}, the defining equation for the cocycles of $s$ modulo $d$ produces a set of descent equations in lower form degrees, and the invariants are distinguished according to whether the descent equations are of zero or nonzero length, corresponding respectively to the \emph{local} and \emph{global} conformal invariants.
The local conformal invariants are (the integral of) scalar densities 
that are strictly Weyl invariant. They can be built using various techniques, be them algebraic \cite{Boulanger:2004zf,Boulanger:2004eh} or geometric 
\cite{curry_gover_2018,Thomas,Thomas1932,Bailey-et-al94,Eastwood96,Gerlach99,GOVER2001,
Alexakis:2005ft,Cap-Slovak09}, for some examples.  
The global invariants are scalar densities that are Weyl invariant 
only up to a total derivative, 
thereby producing a non-trivial descent equations with respect to the 
Weyl part of the BRST differential. 

Our main result, summarised in our Theorem at the end of  Section \ref{consistency}, 
is that the global conformal invariants are further split into 
two distinct subclasses: The (integral of the) Euler density in even dimension 
$n=2m$ and the (integral of) scalar densities built out of the Lorentz Chern--Simons 
forms ${L}^{4p-1}_{CS}$ in all dimensions $n=4p-1\,$, $p\in \mathbb{N}^{*}\,$.
As far as we could see, this latter (infinite) class of global conformal invariants 
had been missed in previous investigations \cite{Alexakis:2005ft,AlexakisBook}. 
As we also show in this note, by taking the Euler-Lagrange derivative 
of ${L}^{4p-1}_{CS}$ with respect to the spacetime metric,  
we obtain a rank-two tensor density in the corresponding space of 
dimension  $n=4p-1\,$ that is \emph{strictly} Weyl invariant, on top 
of being covariantly conserved and traceless. 

Our work, on the one hand generalises the analyses 
of \cite{Deser:1981wh,vanNieuwenhuizen:1985cx,Horne:1988jf} devoted to the 
three-dimensional case $p=1\,$,  
and on the other hand completes the results obtained in \cite{AlexakisBook}, 
where by assumption only the spaces of even dimensions $n=2m$ were considered.

The latter work was stimulated by a conjecture of Deser and Schwimmer 
\cite{Deser:1993yx} related to the classification of Weyl anomalies 
in arbitrary dimensions. 
It is worth stressing that the classification of Weyl anomalies is a different problem 
compared to the classification of global conformal invariants.  
An anomaly in quantum field theory is the nonzero variation of the effective 
(quantum) action under infinitesimal transformations that are symmetries of the classical theory. 
Therefore, an anomaly is linear in  the infinitesimal gauge parameters. 
The most powerful treatment for the quantisation of gauge systems \cite{Henneaux:1992ig} 
is given by the BRST formalism where the gauge parameters are replaced by quantities 
transforming under the symmetry group in the same way as the gauge parameters, 
but having the opposite Grassmann parity. They are called \emph{ghosts}. The 
infinitesimal variations are generated by the action of the BRST differential. 
Since an anomaly is an infinitesimal variation, it obeys an integrability condition 
known as the Wess-Zumino (WZ) consistency condition \cite{Wess:1971yu}, 
which is nothing but a cocycle condition for the BRST differential.
The modern treatment of anomalies in QFT --- see, e.g., \cite{Bertlmann:1996xk} --- 
therefore entails solving  the  WZ consistency condition at ghost number one. 
It turns out that this is a cohomological problem: An anomaly is trivial if it can be redefined 
away by the addition of a local counterterm in the classical action, which translates into the 
fact that trivial anomalies are exact cochains (called coboundaries) in the BRST differential. 
The computation of the cohomology 
$H^{1,n}(s|d)$ of the BRST differential at ghost number \emph{one} and top form degree in the (jet) space of the metric, the ghosts and their derivatives, gives the general solution of the WZ consistency condition for the corresponding anomaly. 
This problem was solved in \cite{Boulanger:2007ab,Boulanger:2007st} for  
quantum systems for which all the classical symmetries (including diffeomorphisms) 
can be preserved at the quantum level, except for Weyl symmetry.
 
As we anticipated, the classification of global conformal invariants is also given 
by the cohomology of the associated BRST differential in top form degree $n\,$, 
but this time, at ghost number \emph{zero}, i.e., $H^{0,n}(s|d)\,$. 
The two cohomological groups $H^{1,n}(s|d)$ and $H^{0,n}(s|d)$ present some similarities 
--- the approach we follow in the current paper closely follows the steps taken 
in \cite{Boulanger:2007ab,Boulanger:2007st} ---, but also striking differences. 
The latter group turns out to be much larger than the former, 
as we shall see, due to the existence of the global conformal invariants given by the Lorentz-Chern-Simons forms ${L}^{4p-1}_{CS}$ 
in all dimension $n=4p-1\,$, $p\in \mathbb{N}^{*}\,$.  
It is a fact that, by multiplying the local Weyl invariants and 
the Euler-Gauss-Bonnet invariants by the Weyl scalar $\phi(x)\,$ 
one produces the type B and the type-A Weyl anomalies, respectively. 
However, one does \emph{not} obtain any consistent Weyl anomaly by 
multiplying the Lorentz-Chern-Simons densities ${L}^{4p-1}_{CS}$ 
by $\phi(x)\,$. 

The plan of the paper is as follows. 
In section \ref{sec:cohosetting} we give the cohomological reformulation of the problem that
consist in giving the general decomposition of global conformal invariants. In section \ref{consistency},
we solve the cohomological problem. In section \ref{sec:EOMLCS} the compute the Euler-Lagrange 
derivative of the Lorentz-Chern-Simons Lagrangians and exhibit the main properties of the resulting 
tensorial density. Finally, we give our conclusions in section \ref{sec:Ccl}.

\section{Cohomological setting}
\label{sec:cohosetting}

For the problem at hand, the corresponding BRST
differential decomposes into $s=s_{\!_D}+s_{\!_W}$, where $s_{\!_D}$ is 
the BRST differential corresponding to the diffeomorphisms and  
$s_{\!_W}$ corresponds to the Weyl transformations. 
Apart from the (invertible) metric, the other fields involved in the problem 
are the diffeomorphisms ghosts $\xi^{\mu}$ and the Weyl ghost $\omega\,$. 
They have a ghost number $\mathrm{gh}(\xi^{\mu})=\mathrm{gh}(\omega)=1\,$. 
Spacetime indices are denoted by Greek letters and run over the values $0,1,\ldots,n-1\,$. 
Flat, tangent space indices are denoted by Latin letters. 
The action of the BRST differential $s$ on the fields 
$\Phi^A=\{g_{\mu\nu},\xi^{\mu},\omega\}$ is given by
\begin{eqnarray}
s_{\!_D} g_{\mu\nu} &=& \xi^{\rho}\partial_{\rho}g_{\mu\nu}+\partial_{\mu}\xi^{\rho}g_{\rho\nu}
+\partial_{\nu}\xi^{\rho}g_{\mu\rho}\;,
\label{sdg} \\
s_{\!_W} g_{\mu\nu} &=& 2\,\omega \,g_{\mu\nu}\,,\qquad 
s_{\!_D} \xi^{\mu} = \xi^{\rho}\partial_{\rho}\xi^{\mu}\;,
\label{sdx} \\
s_{\!_D} \omega &=& \xi^{\rho}\partial_{\rho}\omega\;,
\quad s_{\!_W} \xi^{\mu} = 0 = s_{\!_W} \omega\;. 
\label{sdw} 	
\end{eqnarray}    
The action of $s$ is extended to the derivatives of $\Phi^{A}$ by demanding 
the anticommutativity relation $\{s,d\}=0\,$ with the convention that the ghosts 
$\omega$ and $\xi^{\mu}$ anticommute with $dx^{\mu}\,$.

The conformal invariants are thus solutions of the following cohomological problem
\begin{eqnarray}
	s \,a^{0,n} + d\, a^{1,n-1} = 0\,, ~\quad\quad~a^{0,n} \neq    d\, b^{0,n-1}\;.
\label{WZbis}
\end{eqnarray}
The first equation is the cocycle condition, while  the second one, referred to as 
the coboundary condition, reflects that fact that a conformal invariant is trivial if it 
reduces to the integral over the spacetime manifold of a total derivative. 
In case the manifold has no boundary, such an invariant vanishes, by Stoke's 
theorem.  
All the cochains $a^{0,n}$, $a^{1,n-1}$ and $b^{0,n-1}$ are \emph{local} forms
and $d$ is the total exterior derivative.   
A local $p\,$-form $b^p$ depends on the fields $\Phi^A$ 
and their derivatives up to some finite (but otherwise unspecified) order, 
which is denoted by  
$b^p=\frac{1}{p!}\,d x^{\m_1}\ldots d x^{\m_p}\,b_{\m_1\ldots\m_p}(x,[\Phi^A])\,$. 
Finally, acting with $s$ on the cocycle equation of \eqref{WZbis} 
and using the algebraic Poincar\'e Lemma reviewed in the section 4 of 
\cite{Barnich:2000zw}, one derives a descent equations, 
\begin{eqnarray}
\left. 
\begin{array}{rc}
	s \,a^{0,n} + d\, a^{1,n-1} &= 0 \quad \\
	s \,a^{1,n-1} + d\, a^{2,n-2} &= 0 \quad \\
	 &\vdots \quad\\
	s \,a^{p-1,n-p+1} + d\, a^{p,n-p} &= 0 \quad\\
	s \,a^{p,n-p}  &= 0\quad
\end{array} \right\}
\label{descent}
\end{eqnarray}
that stops either because $p=n$ or because one meets a cocycle 
$a^{p,n-p}$ of $s\,$. 

By decomposing the cocycle condition (\ref{WZbis}) with respect to the 
Weyl-ghost degree, one finds
\begin{equation}
\left\{
\begin{array}{rc}
	s_{\!_D} a^{0,n} + d\, f^{1,n-1} =& 0\;, 
	\\
	s_{\!_W} a^{0,n} + d\, g^{1,n-1} =& 0\;,
\end{array}
\right. ~ 
	a^{0,n} \neq  d\, b^{0,n-1}\;.
	\label{coho2} 
\end{equation}
The ghost degree of $f^{1,n-1}$ is carried by the (derivatives of the) 
diffeomorphism ghosts $\xi^{\mu}\,$, while the ghost degree of $g^{1,n-1}$ 
is carried by the (derivatives of the) Weyl ghost $\omega\,$.  
In words, we have to find the cocycles  of 
the differential $s_{\!_W}\,$ modulo $d\,$, 
in the cohomology of the diffeomorphism-invariant local $n$-forms.
What will considerably facilitate our task, is that the latter cohomology 
class has already been worked out in \cite{Brandt:1989et}, see Eq. (7.43) therein, 
and also in \cite{Barnich:1995ap}. 
If one denotes by $f_{K}:={\rm Tr}(R^{m(K)})\,$,  $K\in\{1,\ldots,r=[n/2]\}\,$,
the characteristic polynomials of the Lorentz algebra $so(1,n-1)\,$ and 
$q^{0}_{K}$ the corresponding Chern-Simons $(2m(K)-1)$-forms
obeying $dq^{0}_{K} = f_{K}\,$, the general solution of the first equation \eqref{coho2}, 
in absence of antifields, decomposes into two main classes \cite{Brandt:1989et,Barnich:1995ap}:
\begin{align}
a^{0,n} &= \sqrt{-g}\,{L}(\nabla,R,g)d^{n}x \label{ScalarType1}\\
&~ ~+\,\sum_{m}\sum_{K:m(K)=m}q^{0}_{K}\frac{\partial}{\partial f_{K}}\,
P_{m}(f_{1},\ldots , f_{r})\;.\label{Dragon}
\end{align}
In this expression, $L$ is a scalar under diffeomorphisms,  
constructed out of the covariant derivatives 
$\nabla_{\alpha_{1}}\ldots\nabla_{\alpha_{m}}R^{\mu}{}_{\nu\rho\sigma}$ 
of the Riemann tensor, where the indices are contracted with the components of the (inverse) metric. 
For the Lorentz algebra $so(1,n-1)\,$, one has $m(K)=2K\,$. 
The second class of terms \eqref{Dragon} in $a^{0,n}$ above therefore only contributes for spacetimes 
of dimensions $n=4p-1\,$, $p\in\mathbb{N}^{*}\,$. 
Indeed, the polynomials $P_{m}$ decompose as 
$P_{m}=\sum_{\boldsymbol{n}}P_{m,\boldsymbol{n}}\,$ where 
$\boldsymbol{n}\in\mathbb{N}^{r}$  and 
 $P_{m,\boldsymbol{n}}=a_{m,\boldsymbol{n}}\prod_{K:m(K)\geqslant m}(f_{K}){}^{n_{K}}\,$, 
 $a_{m,\boldsymbol{n}}\in\mathbb{R}\,$ and $n_{K}$, $K=1,\ldots,r\,$, being 
 the components of $\boldsymbol{n}\,$.
A polynomial $P_{m}$ will contribute to $a^{0,n}$ when  
$2 \sum_{K} n_{K}\,m(K)\equiv \sum_{K} 4K\, n_{K}  = n+1\,$, so that
the form degree of  $q^{0}_{K}\frac{\partial}{\partial f_{K}}P_{m,\boldsymbol{n}}$ is 
indeed $n=4p-1\,$.    
For more details on the notation the adopted and the Lie-algebra cohomology of 
(pseudo)orthogonal algebras, see  the appendix B of \cite{Barnich:1995ap}.
Taking $n=7$ as a definite example, the second class of Lagrangians provides 
the following two candidates:
\begin{align}
&{\rm Tr}(\Gamma d\Gamma +\tfrac{2}{3}\,\Gamma^{3}){\rm Tr}(R^{2})
\equiv L_{CS}^{3}\,{\rm Tr}(R^{2})\; {\rm and}\; L_{CS}^{7}={\rm Tr}(I_{7})\;,
\nonumber \\
&I_{7} = \Gamma (d\Gamma)^{3}+\tfrac{8}{5}(d\Gamma)^{2}\Gamma^{3}
+\tfrac{4}{5}\Gamma (\Gamma d\Gamma)^{2}+2\,\Gamma^{5}d\Gamma 
+\tfrac{4}{7}\Gamma^{7}
,\nonumber
\end{align}
where $\Gamma$ denotes the matrix-valued $1$-form 
$dx^{\mu}\,\Gamma^{\alpha}{}_{\beta\mu}\,$ whose components 
$\Gamma^{\alpha}{}_{\beta\mu}$ are the Christoffel symbols and Tr$(\cdot)$ 
denotes the matrix trace. In particular,  
Tr$R^{2}\equiv R^{\alpha}{}_{\beta}R^{\beta}{}_{\alpha}$
for $R^{\alpha}{}_{\beta}=\tfrac{1}{2}\,dx^{\mu}dx^{\nu}\,R^{\alpha}{}_{\beta\mu\nu}$ 
the curvature $2$-form. The wedge product symbol is omitted throughout this paper. 

Before tackling the problem \eqref{coho2}, 
it is useful to recall \cite{Brandt:1996mh} how on can reformulate the equations 
for the computation of $H^{g,n}(s\vert d)\,$ in slightly different terms. 
One can perform the Stora trick \cite{Sto77} that consists in uniting the differentials 
$s$ and $d$ into a single differential $\tilde{s}=s + d\,$. 
Then, the whole descent equations \eqref{descent} 
and the corresponding coboundary conditions are encapsulated in 
\begin{eqnarray}
	\tilde{s} \,{\alpha} = 0\,, \quad {\alpha}\neq \tilde{s}\,{\zeta}+constant
\label{WZ}
\end{eqnarray}
for the local \emph{total forms} ${\alpha}$ and ${\zeta}$ of total degrees (the sum of the form degree and the ghost number) $G(\alpha)= \mathrm{g} +n$ and 
$G(\zeta)=\mathrm{g}+n-1\,$, respectively, where local total forms are by definition 
formal sums of local forms with different form degrees and ghost numbers, 
${\alpha}=\sum_{p=0}^n a^{G-p,p}\,$.   
As proved in \cite{Brandt:1996mh}, the cohomology of ${s}$ in the space of local 
functionals (integrals of local $n$-forms) and at ghost number $\mathrm{g}$ is locally 
isomorphic to the cohomology of $\tilde{s}$ in the space of local total forms
at total degree~$G=\mathrm{g}+n\,$.
Furthermore, the cohomological problem can be restricted, locally, 
to the $\tilde{s}$-cohomology on local total forms belonging to a subspace 
${\cal{W}}\,$ of the space of local total forms~\cite{Brandt:1996mh}:
\begin{eqnarray}
	&\tilde{s}\,\alpha({\cal{W}}) = 0\,,\quad 
	\alpha({\cal{W}})\neq \tilde{s}\,\zeta({\cal{W}})+constant\,,&
\label{cohoproblem1} \\
	&\quad G(\alpha) (\alpha)=n+\mathrm{g}\,, \quad G(\zeta)=n+\mathrm{g}-1\,.&
\nonumber	
\end{eqnarray}
The subspace ${\cal{W}}$, closed under the action of $\tilde{s}$, 
is given by local total forms depending on tensor fields 
$\{ {\cal{T}}^i\}$ at total degree zero and on generalized (or algebraic \cite{DuboisViolette:1985jb}) connections 
$\{\widetilde{C}^N\}$ at total degree unity. 
The latter decompose into a part with ghost number one and form degree zero 
plus a part having ghost number zero but
form degree unity: $\widetilde{C}^{N}=\widehat{C}^{N}+\ca^{N}\,$. 
For a purely gravitational theory in metric formulation,  
invariant under both diffeomorphisms and Weyl transformations, 
the space ${\cal{W}}$ was identified in~\cite{Boulanger:2004eh}.

As explained in \cite{Brandt:1996mh}, the solution of the 
cohomological problem (\ref{WZ}) will thus have the form   
\begin{eqnarray}
 \alpha({\cal{W}})=\widetilde{C}^{N_1}\ldots\widetilde{C}^{N_n}
 \,a_{N_1\ldots N_{n}}({\cal{T}})\;.
 \nonumber 	
\end{eqnarray}
An invariant at ghost number $\mathrm{g}=0$ will be given 
(up to an unessential constant coefficient) 
by the top form-degree component of the local total form $\alpha({\cal{W}})$:
\begin{eqnarray} 
a^{0,n} = {\cal{A}}^{N_1}\ldots{\cal{A}}^{N_n} \,a_{N_1\ldots N_{n}}
({\cal{T}})\,. 	
\nonumber
\end{eqnarray} 

Now, we are ready to attack the system \eqref{coho2}.  
Our strategy is to solve (\ref{cohoproblem1}) at total degree $G=n$  
with $\tilde{s}$ replaced by $\tilde{s}_{\!_W}={s}_{\!_W}+d\,$ and
taking the first equation of \eqref{coho2} into account, meaning 
that one works in the space of local total forms that are invariant under diffeomorphisms, see \cite{Brandt:1996mh,Barnich:1995ap}.
%
\section{Solution of the consistency condition }
\label{consistency}
%
To reiterate, we must look for $\tilde{s}_{\!_D}$-invariant 
local total forms $\alpha({\cal{W}})$ of total degree $n$ satisfying
\begin{eqnarray}
	&\tilde{s}_{\!_W}\alpha({\cal{W}})=0\,,
	\quad\alpha({\cal{W}})\neq \tilde{s}_{\!_W}\zeta({\cal{W}})+constant\,,&
  \label{cohoproblemweyl}
\end{eqnarray}
where $\zeta({\cal{W}})$ must be $\tilde{s}_{\!_D}$-invariant. 
The solution will take the general form
\begin{eqnarray}
\alpha({\cal{W}})= \widetilde{C}^{N_1}\ldots\widetilde{C}^{N_n}\;
a_{N_1\ldots N_{n}}({\cal{T}})\,.&
\label{ansatz}
\end{eqnarray}
We refer to \cite{Boulanger:2004eh,Boulanger:2007ab,Boulanger:2007st} 
for the detailed explanation of the various symbols that appear in the above equation 
(\ref{ansatz}), in particular, 
the space $\cal T$ of tensors and the generalised connections $\widetilde{C}$
for a classical system  invariant under diffeomorphisms and Weyl rescalings of the 
spacetime metric. 

Very briefly, 
the space ${\cal{T}}$ of tensor fields is generated by the undifferentiated 
metric components $g_{\mu\nu}$ together with the  $W$-tensors 
$\{W_{\Omega_i}\}$, $i\in\mathbb{N}\,$ that are tensors under general coordinate
transformations and transform under $s_{\!_W}$ according to 
$s_{\!_W} W_{\Omega_i}=\omega_{\alpha} \mathbf{\Gamma}^{\alpha}W_{\Omega_i}\,$, where
$\omega_{\alpha}=\partial_{\alpha}\omega\,$ and the $n$ generators $\mathbf{\Gamma}^{\alpha}\,$, 
$\alpha \in \{ 0, \ldots, n-1\}\,$, act only on the $W$-tensors. 
These tensors are built recursively with the help of the formula 
$W_{\Omega_k}=(\nabla_{\alpha_k}+P_{\beta\alpha_k}\mathbf{\Gamma}^{\beta})W_{\Omega_{k-1}}={\cal{D}}_{\alpha_k}W_{\Omega_{k-1}}\,$, 
where $P_{\alpha\beta} = $
$\frac{1}{n-2}\,\Big(R_{\alpha\beta}-\frac{1}{2(n-1)}\,g_{\alpha\beta}R \Big)\,$ 
is the Schouten tensor 
and $W_{\Omega_0}$ denotes the conformally invariant Weyl tensor with three covariant and 
one contravariant indices.
For the components of the various tensors, we use an (arbitrary) 
local coordinate system with basis indices denoted by Greek letters.  
The symbol $\nabla=dx^{\mu}\nabla_{\mu}$ denotes the usual torsion-free metric-compatible covariant 
derivative associated with the Christoffel symbols $\Gamma_{~\,\nu\rho}^{\mu}\,$,   
while ${{R}}_{\alpha\beta}=R^{\mu}_{~\,\alpha\mu\beta}\,$ is the Ricci tensor
with $R^{\mu}_{~\;\nu\rho\sigma}=\partial_{\rho}\Gamma_{~\,\nu\sigma}^{\mu}+\ldots$
the components of the Riemann tensor in some local coordinate system. 
The scalar curvature is given by $R=g^{\alpha\beta}R_{\alpha\beta}\,$.
\noindent The components of the Weyl tensor can be written as
\begin{eqnarray}
W^{\mu}_{~\;\nu\rho\sigma} = R^{\mu}_{~\;\nu\rho\sigma}
-2\left(\delta^{\mu}{}_{\,[\rho}P_{\sigma]\nu}-g_{\nu[\rho}P_{\sigma]}{}^{\mu} \right)\,,	
\end{eqnarray}
where curved (square) brackets denote strength-one total (anti)symmetrization.

We now proceed with (\ref{cohoproblemweyl}) and expand $\alpha({\cal{W}})$ 
in powers of the connection $1$-form ${\Gamma}^{\nu}{}_{\mu}\,$, 
\begin{eqnarray}
	\alpha({\cal{W}})=\sum_{k=0}^{M}\alpha_{k}\;,~
	N_\Gamma\,\alpha_k=k\,\alpha_k \text{ with }
N_\Gamma = {\Gamma}^{\nu}{}_{\mu}\frac{\partial^L}{\partial {\Gamma}^{\nu}{}_{\mu}} \;,
\nonumber
\end{eqnarray}
where $\partial^L$ denotes the partial derivative taken from the left.   
On ${\cal{W}}\,$, the differential $\tilde{s}_{\!_W}$ decomposes into three parts, 
\begin{eqnarray}
	\tilde{s}_{\!_W}\, \alpha({\cal{W}}) = (\tilde{s}_+  + \tilde{s}_0 + \tilde{s}_{-})\;
	\alpha({\cal{W}})
	\nonumber
\end{eqnarray}
that have $N_\Gamma\,$-degrees $1$, $0$, $-1$ respectively. 
The action of $\tilde{s}_+$, $\tilde{s}_{0}$ and 
$\tilde{s}_-$ is summarised in Table \ref{ta}, together with
\begin{eqnarray}
	\tilde{s}_-\Gamma^{\nu}{}_{\mu}&=&\frac{1}{2}\,d{x}^{\rho} 
	d x^{\sigma}W_{~\,\mu\rho\sigma}^{\nu}+
  {\cal{P}}^{\nu\alpha}_{\beta\mu}\;\tilde{\omega}_{\alpha}\;d{x}^{\beta}\,,
\nonumber
\end{eqnarray} 
where 
\begin{align}
{\cal{P}}^{\mu\alpha}_{\rho\nu} &= - g^{\mu\alpha}g_{\rho\nu} +\delta^{\mu}_{\rho}\delta^{\alpha}_{\nu}
  +\delta^{\mu}_{\nu}\delta^{\alpha}_{\rho}\;,\\
 \tilde{\omega}_{\alpha} &= \partial_{\alpha}\omega - dx^{\beta}P_{\beta\alpha}
 = \partial_{\alpha}\omega+{\cal A}_{\alpha}\;. \label{calA}
\end{align}
\begin{table}[h]
\begin{center}
\begin{tabular}{|c||c|c|c|}
\hline
     & $\tilde{s}_+$ &  $\tilde{s}_{0}$  &  $\tilde{s}_-$ \\
\hline \hline
$\Gamma^{\nu}{}_{\mu}$ & $-\Gamma^{\nu}{}_{\alpha}\Gamma^{\alpha}{}_{\mu}$ & 
$0$ &  $\tilde{s}_-\Gamma^{\nu}{}_{\mu}$ \\
\hline
$\tilde{\omega}_{\alpha}$ & $\Gamma^{\beta}{}_{\alpha}\tilde{\omega}_{\beta}$ & 
$\frac{1}{2}\,d x^{\rho} d x^{\sigma} {C}_{\alpha\rho\sigma}$ & $0$ \\
\hline
$\omega$ & $0$ & $dx^{\mu} \tilde{\omega}_{\mu}$ & $0$ \\
\hline
$g_{\mu\nu}$ & $\Gamma^{\beta}{}_{\alpha}\Delta^{\alpha}{}_{\beta}g_{\mu\nu}$ & 
$2\,\omega \,g_{\mu\nu}$ & $0$ \\
\hline
$W_{\Omega_i}$ & $\Gamma^{\beta}{}_{\alpha}\Delta^{\alpha}{}_{\beta}W_{\Omega_i}$ & 
$dx^{\mu}{\cal{D}}_{\mu}W_{\Omega_i}+\tilde{\omega}_{\alpha} \mathbf{\Gamma}^{\alpha}
W_{\Omega_i}$ & $0$ \\
\hline
\end{tabular}
\caption{Decomposition of the action of $\tilde{s}_{\!_W}$\label{ta}, where $\Delta^{\alpha}{}_{\beta}$
are the generators of $Gl(n)\,$.}
\end{center}
\end{table}

The cocycle condition $\tilde{s}_{\!_W}\alpha=0\,$ thus decomposes into 
\begin{eqnarray}
	0 &=& \tilde{s}_+ \alpha_M \label{1} \\
	0 &=& \tilde{s}_{0} \alpha_M + \tilde{s}_+ \alpha_{M-1} \label{2} \\
	0 &=& \tilde{s}_- \alpha_M + \tilde{s}_{0} \alpha_{M-1} + 
	\tilde{s}_+ \alpha_{M-2} 
	\nonumber 
	\\
	&\vdots&\nonumber
\end{eqnarray}
In the first equation, a contribution of the form $\tilde{s}_+ \beta_{M-1}$ 
can be redefined away by subtracting the trivial piece $\tilde{s}_{\!_W}\beta_{M-1}$ 
from $\alpha\,$. The solution of equation (\ref{1}) is known because we know the Lie algebra 
cohomology of $\mathfrak{gl}(n)\,$.  
Indeed, $\mathfrak{gl}(n)\cong \mathbb{R}\oplus \mathfrak{sl}(n)\,$ is reductive.  
Since all the fields of $\cw$ transform according to finite-dimensional linear 
representations of $\mathfrak{gl}(n)$, we have
\begin{eqnarray}
	\alpha_M = \varphi_i(dx, \omega, \tilde{\omega}_{\alpha}, {\cal{T}}) 
	P^i(\tilde{\theta})\,, \quad
	\tilde{s}_+ \varphi_i = 0\,.
	\label{sollie} 
\end{eqnarray}
The $P^i(\tilde{\theta})$ are linearly independent polynomials in the primitive elements 
$\tilde{\theta}_K= \mathrm{Tr}(\Gamma^{2K-1})\,$ of the Lie algebra cohomology of $\mathfrak{gl}(n)$, 
$K=2,3,\ldots,n\,$. The primitive elements correspond to the independent Casimir operators 
of $\mathfrak{gl}(n)\,$. 
Inserting (\ref{sollie}) into (\ref{2}) gives 
$$(\tilde{s}_{0} \varphi_i)P^i(\tilde{\theta})+\tilde{s}_+\alpha_{M-1}=0\,.$$
From this equation and the knowledge of the Lie algebra cohomology of $\mathfrak{gl}(n)\,$, 
we deduce 
\begin{eqnarray}
	\tilde{s}_{0} \varphi_i(dx,\omega,\tilde{\omega}_{\alpha},{\cal{T}})
	=0;\, \quad \forall\; i\,.
	\label{4}
\end{eqnarray}
We can assume that none of the $\varphi_i$'s is of the form 
$\tilde{s}_{\!_W}\vartheta(dx,\omega,\tilde{\omega}_{\alpha},{\cal{T}})$ 
because otherwise we could remove that particular $\varphi_i$ by subtracting the trivial piece 
$\tilde{s}_{\!_W}(\vartheta P^i)$ from $\alpha\,$. 
Such a subtraction does not reintroduce a term 
$\tilde{s}_+\beta_{M-1}$ in (\ref{sollie}) because 
of the definition of the $P^i(\tilde\theta)$'s.    

Hence, since the $\varphi_i$'s do not depend on the $\Gamma^{\nu}{}_{\mu}$'s, 
we see that they are determined by the $\tilde{s}_{\!_W}$-cohomology in the space of 
$\mathfrak{gl}(n)$-invariant local total forms 
$\varphi(dx,\omega,\tilde{\omega}_{\alpha},{\cal{T}})$. 
[The coboundary condition 
$\varphi(dx,\omega,\tilde{\omega}_{\alpha},{\cal{T}})=\tilde{s}_{\!_W} \vartheta(dx,\omega,\tilde{\omega}_{\alpha},{\cal{T}})$
requires $\vartheta$ to be $\mathfrak{gl}(n)$-invariant, 
by expanding the equation in $\Gamma^{\nu}{}_{\mu}$.]
We are thus led to solve 
\begin{eqnarray}
	&\tilde{s}_{\!_W} \varphi(dx,\omega,\tilde{\omega}_{\alpha},{\cal{T}})=0\,,&
	\label{equivcoho1} \\
	&\quad \varphi(dx,\omega,\tilde{\omega}_{\alpha},{\cal{T}})\neq 
	\tilde{s}_{\!_W} \vartheta(dx,\omega,\tilde{\omega}_{\alpha},\ct)\,,&
	\label{equivcoho2} \\ 
	&\tilde{s}_+ \varphi = 0 = \tilde{s}_+\vartheta\,.&  
  \label{equivcoho3}
\end{eqnarray}

In order to solve the above equations, we decompose the relation
$\tilde{s}_{\!_W} \varphi(dx,\omega,\tilde{\omega}_{\alpha},{\cal{T}})=0$ 
into parts with definite degree in the appropriately symmetrized $W$-tensor 
fields (see \cite{Boulanger:2004eh}) 
and analyze it starting from the part with lowest degree. 
The decomposition is unique and thus well-defined thanks to the algebraic independence of 
the appropriately symmetrized $W$-tensors. 
The decomposition of 
$\tilde{s}_{\!_W}$ by this filtration takes the form 
\[
\tilde{s}_{\!_W}=\sum_{k\geqslant 0}\tilde{s}_{\!_W}^{(k)}\, , 
\qquad [N_W,\tilde{s}_{\!_W}^{(k)}]=k\,\tilde{s}_{\!_W}^{(k)}\;,
\]
where $N_W$ is the counting operator for the 
--- appropriately symmetrized --- $W$-tensors. 

The $\mathfrak{gl}(n)$-invariant local total form 
$\varphi(dx,\omega,\tilde{\omega}_{\alpha},{\cal{T}})$ 
decomposes into a sum of $\mathfrak{gl}(n)$-invariant terms 
\begin{align*}
	\varphi(dx,\omega,\tilde{\omega}_{\alpha},{\cal{T}}) &=
	\varphi_{(0)}(dx,\omega,\tilde{\omega}_{\alpha},g_{\mu\nu}) 
	\nonumber \\ 
	&\qquad +\;
	\sum_{k>0}\varphi_{(k)}(dx,\omega,\tilde{\omega}_{\alpha},{\cal{T}})\,,\\
	N_W\,\varphi_{(k)} &= k \,\varphi_{(k)}\,.
\end{align*}  
The condition $\tilde{s}_{\!_W}\varphi=0$ requires, at lowest order in the tensor fields, 
\begin{eqnarray}
	\tilde{s}_{\!_W}^{(0)} \varphi_{(0)}(dx,\omega,\tilde{\omega}_{\alpha},g_{\mu\nu})=0\,.
	\label{cocyclegammazero}
\end{eqnarray}
Furthermore, we can remove any piece of 
the form 
$\tilde{s}_{\!_W}^{(0)} \vartheta_{(0)}(dx,\omega,\tilde{\omega}_{\alpha},g_{\mu\nu})$ 
from $\varphi_{(0)}$ by subtracting the trivial piece $\tilde{s}_{\!_W}\vartheta_{(0)}$ 
from $\varphi\,$. 
Hence, $\varphi_{(0)}$ is actually determined 
by the $\tilde{s}_{\!_W}^{(0)}$-cohomology in the space of $\mathfrak{gl}(n)$-invariant local total forms with no dependence on the $W$-tensors.
The action of $\tilde{s}_{\!_W}^{(0)}$ on the various variables
$\{ dx^{\mu},\omega,\tilde{\omega}_{\alpha},g_{\mu\nu}\}\,$ is as follows. 
First, $\tilde{s}_{\!_W}^{(0)}\tilde{\omega}_{\alpha}=\Gamma^{\beta}{}_{\alpha}\,\tilde{\omega}_{\beta}\,$
and  
$\tilde{s}_{\!_W}^{(0)}dx^{\mu}= - \Gamma^{\mu}{}_{\nu}\,dx^{\nu}\equiv 0\,$, which coincide with 
the action of $\mathfrak{gl}(n)$ on them. The action on $dx^{\mu}$ is identically zero because the
Levi-Civita connection has zero torsion.
Then, $\tilde{s}_{\!_W}^{(0)}\omega = d\omega$ is 
the exterior derivative on $\omega\,$, and finally
$\tilde{s}_{\!_W}^{(0)}g_{\mu\nu} = \Gamma^{\alpha}{}_{\beta}\Delta^{\beta}{}_{\alpha}\,g_{\mu\nu}+
2\omega g_{\mu\nu}\,$ where 
$\Gamma^{\alpha}{}_{\beta}\Delta^{\beta}{}_{\alpha}\,g_{\mu\nu}=\Gamma^{\alpha}{}_{\mu}\,g_{\alpha\nu}+
\Gamma^{\alpha}{}_{\nu}\,g_{\mu\alpha}\,$.
The  contributions  
$\Gamma^{\beta}{}_{\alpha}\,\tilde{\omega}_{\beta}\,$ and  
$\Gamma^{\alpha}{}_{\beta}\Delta^{\beta}{}_{\alpha}g_{\mu\nu}$
give a zero net result inside $\varphi_{(0)}$ because of the condition that 
$\varphi_{(0)}(dx,\omega,\tilde{\omega}_{\alpha},g_{\mu\nu})$ is
$\mathfrak{gl}(n)$-invariant. 

Turning to the co-boundary condition \eqref{equivcoho2} on $\varphi_{(0)}\,$, we can assume 
$\varphi_{(0)}\neq \tilde{s}_{\!_W}^{(0)} \vartheta_{(0)}(dx,\omega,\tilde{\omega}_{\alpha},g_{\m\n})\,$.
We decompose  $\varphi_{(0)}= \omega \ell_{(0)}(dx,\tilde{\omega}_{\alpha},g_{\mu\nu})
+m_{(0)}(dx,\tilde{\omega}_{\alpha},g_{\mu\nu})\,$, where 
\begin{align}
\ell_{(0)} = & \sum_{p=0}^{n} \Big(\eta_{p} (dx^{\alpha}\tilde{\omega}_{\alpha})^{p}
+ \sigma_{p} \sqrt{-g}\, \varepsilon_{\mu_{1}\ldots\mu_{p}\nu_{1}\ldots\nu_{n-p}}
\nonumber \\
& dx^{\mu_{1}} \ldots dx^{\mu_{p}}  
g^{\nu_{1}\alpha_{1}} \ldots
g^{\nu_{n-p}\alpha_{n-p}} \, \tilde{\omega}_{\alpha_{1}} \ldots \tilde{\omega}_{\alpha_{n-p}}
   \Big)\nonumber\;.
\end{align}
The quantity $m_{(0)}$ has the same expression as $\ell_{(0)}$, 
except that the constants $\eta_{p}\,$ and $\sigma_{p}$ are replaced by constants 
$\kappa_{p}\,$ and $\zeta_{p}\,$, respectively.
Actually, the coboundary condition on $\varphi_{(0)}$ allows us to set all the 
$\kappa_{p}\,$'s to zero, since $dx^{\alpha}\tilde{\omega}_{\alpha}\equiv d\omega$ and 
$(d\omega)^{p}= \tilde{s}_{\!_W}^{(0)} [\omega (d\omega)^{p-1}]\,$.
As for the terms proportional to $\sigma_{p}$ in $\varphi_{(0)}\,$, all but one of them 
are $\tilde{s}_{\!_W}^{(0)}\,$-exact. The only non-trivial term is the one for which $p=n/2\,$, 
on account of the following equality:
\begin{align}
\tilde{s}_{\!_W}^{(0)}\Big( & \sqrt{-g}\, \varepsilon_{\mu_{1}\ldots\mu_{p}\nu_{1}\ldots\nu_{n-p}}
dx^{\mu_{1}} \ldots dx^{\mu_{p}} \nonumber \\
& \qquad g^{\nu_{1}\alpha_{1}} \ldots
g^{\nu_{n-p}\alpha_{n-p}} \, \tilde{\omega}_{\alpha_{1}} \ldots \tilde{\omega}_{\alpha_{n-p}}
\Big) \nonumber \\
& = (2p-n)\omega \Big( \sqrt{-g}\, \varepsilon_{\mu_{1}\ldots\mu_{p}\nu_{1}\ldots\nu_{n-p}}
dx^{\mu_{1}} \ldots dx^{\mu_{p}} 
\nonumber \\
& \qquad g^{\nu_{1}\alpha_{1}} \ldots g^{\nu_{n-p}\alpha_{n-p}}\tilde{\omega}_{\alpha_{1}} \ldots \tilde{\omega}_{\alpha_{n-p}}\Big)\;.\nonumber
\end{align}
Now, the cocycle condition (\ref{cocyclegammazero}) 
gives
\[
dx^{\m}\tilde{\omega}_{\m}\ell_{(0)} + \tilde{s}_{\!_W}^{(0)} m_{(0)}=0\,
\]
and a straightforward computation shows that the only possibility for it to vanish 
is that all the constants $\zeta_{p}$ should be zero except for $\zeta_{p}$ with 
$p=n/2\,$, therefore enforcing the spacetime dimension to be even, $n=2m\,$. 
It also requires that all coefficients $\eta_{p}$ should vanish. 

As a result, the most general non-trivial solution of 
\eqref{cocyclegammazero} for $\varphi_{(0)}$
in the cohomology of $\tilde{s}_{\!_W}^{(0)}$ is 
\begin{align}
	\varphi_{(0)} & = (\sigma\,\omega+\eta)\sqrt{-g}\,
g^{\nu_{1}\alpha_{1}}\ldots g^{\nu_{m}\alpha_{m}}
\varepsilon_{\alpha_{1}\ldots\alpha_{m}\mu_1\ldots\mu_m}
\nonumber \\
	& dx^{\mu_1}\ldots dx^{\mu_m}\,\tilde{\omega}_{\nu_1}\ldots\tilde{\omega}_{\nu_m}\;,
	\quad n=2m\;.
  \nonumber
\end{align}
The contribution proportional to the constant $\sigma$ gives rise to the type-A Weyl anomaly studied in \cite{Boulanger:2007ab}, therefore is not to be considered for a candidate conformal invariant, at ghost number zero. We are therefore left with 
\begin{align}
	\varphi_{(0)} & = \sqrt{-g}\,
g^{\nu_{1}\alpha_{1}}\ldots g^{\nu_{m}\alpha_{m}}
\varepsilon_{\alpha_{1}\ldots\alpha_{m}\mu_1\ldots\mu_m}
\nonumber \\
	& dx^{\mu_1}\ldots dx^{\mu_m}\,\tilde{\omega}_{\nu_1}\ldots\tilde{\omega}_{\nu_m}\;,
	\quad n=2m\;,
  \label{fzero} 
\end{align}
up to an irrelevant overall constant coefficient. 

One may now ask what is the completion $\varphi=\varphi_{(0)}+\sum_k \varphi_{(k)}$ 
of (\ref{fzero}) that would be invariant under the full differential 
$\tilde{s}_{\!_W}\,$. 
This can be answered by using a decomposition of $\varphi$ and 
$\tilde{s}_{\!_W}$ with respect to the $\tilde{\omega}_{\alpha}$-degree.  
The differential $\tilde{s}_{\!_W}$ decomposes into a part noted 
$\tilde{s}_{\flat}$ which lowers the 
$\tilde{\omega}_{\alpha}$-degree by one unit, 
a part noted $\tilde{s}_{\natural}$ which does not 
change the $\tilde{\omega}_{\alpha}$-degree and a part noted 
$\tilde{s}_{\sharp}$ which raises the 
$\tilde{\omega}_{\alpha}$-degree by one unit:  
$\tilde{s}_{\!_W}= 
\tilde{s}_{\flat}+\tilde{s}_{\natural}+\tilde{s}_{\sharp}$.
The action of these three parts of $\tilde{s}_{\!_W}$ is given in Table \ref{ta2}. 
\begin{table}[!hbtp]
\begin{center}
\begin{tabular}{|c||c|c|c|}
\hline
     & $\tilde{s}_{\flat}$ &  $\tilde{s}_{\natural}$  &  
     $\tilde{s}_{\sharp}$ \\
\hline \hline
$\tilde{\omega}_{\alpha}$ & 
$\frac{1}{2}\,d x^{\rho} d x^{\sigma} {C}_{\alpha\rho\sigma}$ & 
$\Gamma^{\beta}{}_{\alpha} \tilde{\omega}_{\beta}$ & $0$ \\
\hline
$\omega$ & $0$ & $ 0 $ & $dx^{\mu} \tilde{\omega}_{\mu}$ \\
\hline
$W_{\Omega_i}$ & $ 0 $ & 
$\Gamma^{\beta}{}_{\alpha}\Delta^{\alpha}{}_{\beta}W_{\Omega_i}
+dx^{\mu}{\cal{D}}_{\mu}W_{\Omega_i}$ & 
$\tilde{\omega}_{\alpha} \mathbf{\Gamma}^{\alpha}W_{\Omega_i}$ \\
\hline
$g_{\mu\nu}$ & $ 0 $ & 
$\Gamma^{\beta}{}_{\alpha}\Delta^{\alpha}{}_{\beta}\,g_{\mu\nu}
+2\,\omega\, g_{\mu\nu}$ & $0$ \\
\hline
$\Gamma^{\nu}{}_{\mu}$ & 0 & 
$-\Gamma^{\nu}{}_{\alpha}\Gamma^{\alpha}{}_{\mu}+\frac{1}{2}\,d{x}^{\rho} 
d x^{\sigma}W_{~\,\mu\rho\sigma}^{\nu}$ & 
  ${\cal{P}}^{\nu\alpha}_{\beta\mu}\;\tilde{\omega}_{\alpha}\;d{x}^{\beta}$ \\
\hline
\end{tabular}
\caption{Action of $\tilde{s}_{\!_W}$, decomposed w.r.t the $\tilde{\omega}_{\alpha}$-degree \label{ta2}}
\end{center}
\end{table}

The decomposition of $\varphi$ with respect to the $\tilde{\omega}_{\alpha}$-degree 
reads
\begin{eqnarray}
	\varphi&=&\Phi^{[m]}_m + \Phi^{[m+1]}_{m-1}+\ldots +\Phi^{[n-1]}_1 + \Phi^{[n]}_0\,, 
\nonumber \\
  &&\qquad \Phi^{[m]}_m = \varphi_{(0)}\,,\qquad m=\frac{n}{2}\;, 
\nonumber
\end{eqnarray}
where each term $\Phi^{[n-r]}_r\,$ ($0\leqslant r\leqslant m$) is $\mathfrak{gl}(n)$-invariant, 
possesses a $\tilde{\omega}_{\alpha}$-degree $r\,$ and explicitly contains the 
product of $(n-r)$ $dx$'s. [Some $dx$'s are also contained inside 
the $\tilde{\omega}_{\a}$'s and the Weyl $2$-forms, see Lemma 1 below.]   

Decomposing the cocycle condition $\tilde{s}_{\!_W} \varphi = 0 $ with respect to the 
$\tilde{\omega}_{\alpha}$-degree 
yields the following descent equations
\begin{eqnarray}
   \tilde{s}_{\flat}\Phi^{[n-1]}_1 + \tilde{s}_{\natural} \Phi^{[n]}_0 &=& 0\quad,
\nonumber \\
	\tilde{s}_{\flat}\Phi^{[n-2]}_2 + \tilde{s}_{\natural}\Phi^{[n-1]}_1 + \tilde{s}_{\sharp}\Phi^{[n]}_0 &=& 0 \quad ,
\nonumber \\
            &\vdots&
\nonumber \\
	\tilde{s}_{\flat}\Phi^{[m]}_m + \tilde{s}_{\natural}\Phi^{[m+1]}_{m-1} 
	+ \tilde{s}_{\sharp}\Phi^{[m+2]}_{m-2} &=& 0 \quad ,
\nonumber \\            
  \tilde{s}_{\natural}\Phi^{[m]}_m + \tilde{s}_{\sharp}\Phi^{[m+1]}_{m-1} &=& 0 \quad ,
\nonumber \\
 \tilde{s}_{\sharp}\Phi^{[m]}_m &=& 0 \quad .
\nonumber
\end{eqnarray}

In the following Lemma 1, we give the expression for $\Phi^{[n-r]}_r\,$, 
$r\in\{0,\ldots, m\}\,$, such that $\varphi=\sum_{r=0}^{m}\Phi^{[n-r]}_r$ 
is a solution of $\tilde{s}_{\!_W} \varphi=0\,$ with 
$\Phi^{[m]}_m = \varphi_{(0)}$ given in (\ref{fzero}).
Furthermore, the $n$-form $\Phi^{[n]}_0\,$ is separately 
$\tilde{s}_{\!_W}$-invariant and the top form degree component of  
$\varphi$ 
 is nothing but the Euler density in the spacetime of dimension $n=2m\,$. 
 The scalar density $\beta=\Phi^{[n]}_0$ gives rise to a trivial descent, 
i.e., it is a local conformal  invariant given by contractions of products 
of $m=n/2$ Weyl tensors.

\vspace*{.3cm}

\noindent
{\underline{\it{Lemma 1\,:}}} ~
Let $\psi_{\m_1\ldots\m_{2p}}$ be the local total form 
\begin{eqnarray}
	\psi_{\mu_1\ldots\mu_{2p}} &=& 
	\frac{1}{\sqrt{-g}} \;
	\varepsilon^{\alpha_1\ldots\alpha_r}_{\quad\quad~\,\nu_1\ldots\nu_r\mu_1\ldots\mu_{2p}}
	\nonumber \\
	&& \qquad\qquad \times ~ \tilde{\omega}_{\alpha_1}\ldots\tilde{\omega}_{\alpha_r}
	\;d x^{\nu_1}\ldots d x^{\nu_r}\,,
	\nonumber \\
	p&=&m-r\,,\quad m=n/2\,,\quad r\in\{0,\ldots, m\}\, ,
  \nonumber
\end{eqnarray}
and $W^{\mu\nu}$ the tensor-valued $2$-form
\begin{eqnarray}
	W^{\mu\nu} &=& W^{\mu}_{~\;\lambda}\,g^{\lambda\nu}=
	\frac{1}{2}\,d{x}^{\rho} d x^{\sigma}W_{~\,\lambda\rho\sigma}^{\mu}\,g^{\lambda\nu}\,.
\nonumber
\end{eqnarray}
Then, the local total forms
\begin{eqnarray}
	\Phi^{[n-r]}_{r} = \frac{(-1)^p}{2^p}\,\frac{m!}{r!\,p!}\;\psi_{\m_1\ldots\m_{2p}}\,
	W^{\m_1\m_2}\ldots \,W^{\m_{2p-\!1}\m_{2p}}
\nonumber
\end{eqnarray}
obey the descent equations
\begin{eqnarray}
&& \left\{
\begin{array}{cl}
	\tilde{s}_{\flat}\Phi^{[n-r]}_{r} + \tilde{s}_{\natural}\Phi^{[n-r+1]}_{r-1} &=\; 0\quad,
	\nonumber \\
	\tilde{s}_{\sharp}\Phi^{[n-r]}_{r} &=\; 0\quad,\quad r\in\{1,\ldots, m\}\,,
	\nonumber 
\end{array}\right.
\\
  &&\tilde{s}_{\flat}\Phi^{[n-1]}_{1} \;=\; 0 \;=\; \tilde{s}_{\!_W} \Phi^{[n]}_{0}\;,	
\nonumber
\end{eqnarray}
so that the following relations hold: 
\begin{eqnarray}
	\tilde{s}_{\!_W} {\alpha} \,= \!&0&\! =\; \tilde{s}_{\!_W}{\beta} \;,
	\nonumber\\
	{\alpha}&=&\sum_{r=1}^{m}\Phi^{[n-r]}_{r} \;, \quad {\beta} = \Phi^{[n]}_{0}\,. 
\nonumber
\end{eqnarray}
\vspace*{.1cm}

\noindent
{\underline{\it{Proof\,:}}} ~ The proof follows by direct computation, 
using the tracelessness of the Weyl tensor and with the help of the identity 
$\nabla W^{\alpha\beta}\equiv 2\,{C}_{\gamma}\,g^{\gamma[\alpha}dx^{\beta]}\,$
relating the covariant differential of the Weyl $2$-form $W^{\alpha\beta}$
to the Cotton $2$-form 
${C}_{\alpha}=\frac{1}{2}\,dx^{\mu}dx^{\nu}\,{C}_{\alpha\mu\nu}$. 

\vspace*{.3cm}

\noindent Finally, we have the
\vspace*{.3cm}

\noindent
{\underline{\it{Lemma 2\,:}}} ~
The top form-degree component $a^{0,n}$ of ${\alpha}$ 
in Lemma 1 
satisfies the cocycle condition for the conformal invariants. 
It gives rise to a non-trivial descent in $H(s_{W}\vert d)\,$. 
The invariant $\beta=\Phi^{[n]}_{0}$ satisfies a trivial descent and is 
obtained by taking contractions of products of Weyl tensors 
($m$ of them in dimension $n=2m$). 
The top form-degree component $e^{0,n}$ of $\alpha+\beta$
is proportional to the Euler density of the manifold ${\cal{M}}_{n}\,$. 
Explicitly, 
\begin{align}
e^{0,n} &=\;\tfrac{(-1)^m}{2^m}\sqrt{-g}\,\varepsilon_{\alpha_1\beta_1\ldots \,\alpha_m \beta_m}
\, (R^{\alpha_1 \beta_1}\wedge\ldots\wedge R^{\alpha_m \beta_m})\;.\nonumber \\
& \label{Euler}	
\end{align}
It is the \emph{only} conformal invariant of the type \eqref{ScalarType1} that satisfies a non-trivial 
descent in $H(s_{W}\vert d)\,$, up to the addition of invariants that satisfy a trivial descent. 
\vspace*{.2cm}

\noindent
{\underline{{\it{Proof\,:}}} 
When computing the solutions of (\ref{equivcoho1})--(\ref{equivcoho3}), 
we used an expansion of $\varphi(dx,\omega,\tilde{\omega}_{\alpha},{\cal{T}})$ 
in the number of 
(appropriately symmetrized) $W$-tensors and found a solution starting with a 
$W$-independent term $\varphi_{(0)}\,$ given in (\ref{fzero}). This term, as
we showed, gives rise to (a representative of) the Euler density. 
However, in order to compute the general solutions of 
(\ref{equivcoho1})--(\ref{equivcoho3}), we must determine whether other
solutions exist, that would start with a term $\varphi_{(\ell)}$ with $\ell>0\,$. 
If one returns to the decomposition of local total forms in terms of 
form degree and ghost number, writing 
$\varphi(dx,\omega,\tilde{\omega}_{\alpha},{\cal{T}})=\sum_{r=1}^{q+1} b^{r,p-r+1}$, 
the problem (\ref{equivcoho1})--(\ref{equivcoho3}) 
takes on the usual descent-equation form
\begin{eqnarray}
	s_{\!_W} b^{0,p} + d \,b^{1,p-1} &=& 0\quad,
	\label{ty} \\
	s_{\!_W} b^{1,p-1} + d \,b^{2,p-2} &=& 0\quad,
  \nonumber \\
	&\vdots&
	\nonumber \\
	s_{\!_W} b^{q-1,p-q+1} + d \,b^{q,p-q} &=& 0\quad,
\label{sec}\\
	s_{\!_W} b^{q,p-q}  &=& 0\quad (0\leqslant q\leqslant p \leqslant n), 
\label{first}	
\end{eqnarray}
where every element $b^{i,p-i}$
$(0\leqslant i\leqslant q)$ transforms as a local $(p-i)$-form under 
spacetime diffeomorphisms, so that $d \,b^{i,p-i} = \nabla b^{i,p-i}$
where $\nabla=dx^{\mu}\nabla_{\mu}\,$.  
One assumes that the descent is displayed in its shortest expansion, \textit{i.e.}
that $q$ is minimal. This means that $b^{q,p-q}$ is non-trivial
in $H^{q,p-q}(s_{\!_W}\vert d)$ since otherwise 
$b^{q,p-q}=s_{\!_W} \mu^{q-1,p-q} + d \,\mu^{q,p-q-1}$ and 
(\ref{sec}) would then become $s_{\!_W} [ b^{q-1,p-q+1} - d \mu^{q-1,p-q}] = 0$, 
which, upon redefining $b^{q-1,p-q+1}\,$, 
would imply that the descent has shortened by one step, 
contrary to the shortest-descent hypothesis.
 
A priori, the top  of the descent, $b^{0,p}\,$, possesses a form degree $p\leqslant n$ 
because candidate conformal invariants are obtained by completing 
[see Eqs. (\ref{1})--(\ref{sollie})] 
the product $\varphi(dx,\omega,\tilde{\omega}_{\alpha},{\cal{T}})P(\tilde{\theta})$,   
where $P(\tilde{\theta})$  carries
a non-vanishing form degree, except for the trivial element $P(\tilde{\theta})=1\,$.
As explained in section \ref{sec:cohosetting} and proved 
in \cite{Brandt:1989et,Barnich:1995ap}, 
in the absence of antifields, the general solution of the first equation 
\eqref{coho2} admits only two kinds of terms, summarised in 
\eqref{ScalarType1} and \eqref{Dragon}.  
The second class of terms \eqref{Dragon} consists of the 
Lorentz-Chern-Simons densities, possibly multiplied by characteristic polynomials, 
and therefore contains the connection 
$1$-form $\Gamma^{\nu}{}_{\mu}$ via a non-trivial descent associated with the 
non-trivial polynomials $P(\tilde{\theta})\,$.

We will treat the case of non-trivial $P(\tilde{\theta})\,$'s in 
Lemma~3 and 
pursue the proof of Lemma 2 with $P(\tilde{\theta})=1\,$.
Taking $P(\tilde{\theta})=1\,$ implies that 
one can set $p=n$ in the descent (\ref{ty})--(\ref{first}), without loss of generality. 

The case where $q=0$ means that the descent is trivial and the candidate
conformal invariants have to satisfy $s_{\!_W} a^n_1 = 0\,$. 
These give the strictly Weyl-invariant densities, also called local conformal invariants.
They can be classified and computed systematically along the lines of
\cite{Boulanger:2004zf,Boulanger:2004eh}, for example, or more geometrically, 
using tractor calculus \cite{curry_gover_2018}.

The bottom of the descent is obtained from  
$\alpha({\cal{W}})$ by taking its maximal $\tilde{\omega}_{\alpha}$-degree 
component and taking only the contribution $\omega_{\alpha}$
of $\tilde{\omega}_{\alpha}=\omega_{\alpha}-dx^{\mu}P_{\mu\alpha}$. 
In other words, the bottom of the descent must not depend on 
the $1$-form potential ${\cal{A}}_{\alpha}:=-dx^{\mu}P_{\mu\alpha}\,$, see \eqref{calA}.   
A priori, when determining the most general non-trivial bottom
$b^{q,n-q}$ in (\ref{first}), the dependence on the 
space of $W$-tensors can be complicated. 
However, it was proved in~\cite{Barkallil:2002fp} that, for
any given (super) Lie algebra ${\mathfrak{g}}$, the solutions of non-trivial descents as in 
(\ref{ty})--(\ref{first}) can be computed, without loss
of generality, in the small algebra ${\cal{B}}$ generated by the $1$-form potentials, 
the curvature $2$-forms, the ghosts and the exterior derivative of the ghosts. 

The rest of the proof of Lemma 2 follows then exactly the same lines as in 
the proof of Theorem 2 in \cite{Boulanger:2007st}, 
except that one must strip off the factors $\omega$ appearing in all 
expressions therein. The outcome is that the only possibility for the bottom of the 
descent is given by 
\begin{eqnarray}
b^{m,m}&=& \sqrt{-g}\;\varepsilon_{\sigma_1\ldots\sigma_m\rho_1\ldots\rho_{m}}\;
g^{\sigma_1\tau_1}\ldots g^{\sigma_{m}\tau_m}\;
\nonumber \\
&&\quad\times\;\omega_{\tau_1}\ldots\,\omega_{\tau_m}\,dx^{\rho_1}\ldots\, dx^{\rho_{m}}\;,
 \label{fre}
\end{eqnarray}
which is precisely contained in (\ref{fzero}). The latter term gives rise to the
conformal invariant $\alpha$ presented in  Lemma~1. Because (\ref{fre}) 
is non-trivial in the cohomology $H(s_{\!_W})\,$, 
so is the corresponding $a^{0,n}$ in $H(s_{\!_W}\vert d)\,$.

It is possible to go up all the steps from the bottom \eqref{fre} of the descent 
in $H(s_{\!_W}\vert d)\,$ up to $b^{0,n}$ in form degree $n\,$.
For this, it suffices to use that 
\begin{align}
s_{\!_W}\,R^{\alpha}{}_{\beta} = \nabla\omega^{\alpha}\,dx_{\beta} 
- \nabla\omega_{\beta}\,dx^{\alpha} \label{gammaR}\;,
\end{align}
as well as  $d\,b^{m,m}\equiv \nabla b^{m,m}\,$, which holds true because 
$b^{m,m,}$ is a singlet under $\mathfrak{gl}(n)\,$ and $\nabla dx^{\mu}\equiv 0$ from 
the zero-torsion condition on the connection $\Gamma\,$. 
A direct computation gives
\begin{align}
d\,b^{m,m}  \,+\,& s_{\!_W} b^{m-1,m+1} = 0\,,
\nonumber \\
b^{m-1,m+1} =& -\tfrac{m}{2}\,\sqrt{-g}\;
\varepsilon_{\alpha_1\beta_{1}\ldots\alpha_{m}\beta_{m}}\;
\nonumber \\
&\qquad \times R^{\alpha_{1}\beta_{1}}
\omega^{\alpha_{2}}dx^{\beta_{2}}\ldots \omega^{\alpha_{m}}dx^{\beta_{m}}\;.
\end{align}
Proceeding in the same way, using that $\nabla b^{m-1,m+1}\equiv d \,b^{m-1,m+1}$
as well as \eqref{gammaR}, we readily obtain the whole chain in $H(s_{\!_W}\vert d)\,$:
\begin{align}
& s_{\!_W} b^{m-r-1,m+r+1} + d\,b^{m-r,m+r}   = 0\,, \quad 0\leqslant r \leqslant m-1 \;,
\nonumber \\
& \hspace{3.55cm} s_{\!_W} b^{m,m} = 0\,,\qquad {\rm where}\nonumber \\
& b^{m-r-1,m+r+1} = \frac{(-1)^{r+1}}{2^{r+1}} \,{{m}\choose{r+1}} \,\sqrt{-g}\;
\varepsilon_{\alpha_1\beta_{1}\ldots\alpha_{m}\beta_{m}}\;\nonumber \\
& \hspace{1.2cm} \times R^{\alpha_{1}\beta_{1}}\ldots R^{\alpha_{r+1}\beta_{r+1}} \,
\omega^{\alpha_{r+2}}dx^{\beta_{r+2}}\ldots \omega^{\alpha_{m}}dx^{\beta_{m}}\;.
\nonumber
\end{align}
The top form-degree element $b^{0,n}\,$, obtained for $r=m-1$ above, is 
exactly given by $e^{0,n}$ in \eqref{Euler}, which achieves  the proof of Lemma 2. 
\vspace{.3cm}

Now that we have classified the general structure for the conformal invariants of type 
\eqref{ScalarType1}, we turn to the determination of the possible global conformal invariants 
of type~\eqref{Dragon}, i.e., 
scalar densities that depend on the connection 
$\Gamma$ through the Lorentz-Chern-Simons $4p-1$-forms 
$L_{CS}^{4p-1}\,$, $p\in\mathbb{N}^*$. 
It turns out that \emph{all} the scalar densities of type~\eqref{Dragon} are 
global conformal invariants. 

\newpage

\noindent
{\underline{\it{Lemma 3\,:}}} ~
Let  $\alpha^{4p-1}_{[2m-1]}$ be the total $(4p-1)$-form of 
degree $2m-1$ in the connection $1$-form $\Gamma\,$ defined by
\begin{align}
\alpha^{4p-1}_{[2m-1]} &:= -\tfrac{1}{2m-1}\,{\rm Tr}\left( [\omega dx - R]^{2p-m} \Gamma^{2m-1}\right)\;,
\\ &\hskip 3.5cm m=1,2,\ldots 2p\;,
\nonumber \\
\alpha^{4p-1}_{[0]} &:= 2\omega(d\omega)^{2p-1}\;,
\end{align}
where 
$[\omega dx - R]$ stands for the matrix-valued  total $2$-form with components 
$\omega^{\alpha} dx_{\beta} - R^{\alpha}{}_{\beta}\,$ and $\Gamma$ denotes the matrix-valued $1$-form with $\Gamma^{\alpha}{}_{\beta}\,$ as $1$-form components.
 
These quantities obey the following descent equations:
\begin{align}
\tilde{s}_{-} \alpha^{4p-1}_{[2m-1]} +\tilde{s}_{+} \alpha^{4p-1}_{[2m-3]} &= 0\;,~~ 
m\in \{2, 3, \ldots, 2p\} \;,\\
\tilde{s}_{-} \alpha^{4p-1}_{[1]} +\tilde{s}_{0} \alpha^{4p-1}_{[0]} &= {\rm Tr}R^{2p}\;,\\
 \tilde{s}_{0} \alpha^{4p-1}_{[2m-1]} &= 0\;, \quad \forall~ m\;.
\end{align} 

\noindent
{\underline{{\it{Proof\,:}}} The proof follows by direct computation.

\vspace{.3cm}

As a result of Lemma 3, the following total form 
\begin{align}
\tilde{\alpha}^{4p-1} := \alpha^{4p-1}_{[0]} + \sum_{m=1}^{2p} \alpha^{4p-1}_{[2m-1]}
\end{align}
obeys the equation 
\begin{align}
\tilde{s}_{W} \tilde{\alpha}^{4p-1} = {\rm{Tr}}R^{2p}\;.
\end{align}
By decomposing the latter equation with respect to the form degree, starting from the 
highest form degree to the lowest, one obtains, in a spacetime of dimension $n=4p-1$, 
the following descent equations:
\begin{align}
 d L^{n}_{CS} &= {\rm{Tr}}R^{2p}\;,\\
s_{W}L^{n}_{CS} + da^{1,n-1} &= 0\;,\\
s_W a^{1,n-1} + da^{2,n-2} &= 0\;, \label{ano}\\
& ~~\vdots \nonumber \\
s_{W} a^{2p-1,2p} + d a^{2p,2p-1}&= 0\;,\\
s_{W} a^{2p,2p-1}&= 0\;,\quad a^{2p,2p-1}\equiv \alpha^{4p-1}_{[0]}\;.
\end{align}
Note that equation \eqref{ano} is the Wess-Zumino consistency condition for a Weyl 
(or conformal) anomaly in a submanifold of co-dimension one with respect to 
${\cal M}_{4p-1}\,$.
The consistent Weyl anomaly is the integral, over this co-dimension one manifold 
${\cal M}_{4p-2}\,$, of the $4p-2\,$-form $a^{1,n-1}\,$, where the latter is found to be
\begin{align*}
a^{1,n-1} &= \sum_{m=1}^{2p} \tfrac{(-1)^{m}}{2m-1}\,dx^{\mu}g_{\mu\alpha}
[\Gamma^{2m-1} R^{2p-m-1}]^{\alpha}{}_{\beta}\,g^{\beta\sigma}\omega_{\sigma} \;.
\end{align*}

Finally, we treat the case of the invariants of type \eqref{Dragon} that are not by themselves 
Lorentz-Chern-Simons $(4p-1)$-forms, but the wedge product of these forms with characteristic polynomials $f_{K}={\rm Tr}(R^{m(K)})\,$.
Since the latter $f_{K}$'s are $d$-closed and $s_{W}$-closed, the descent associated with a product of the type $L^{4p-1}_{CS}f_{K_{1}}\ldots f_{K_{m}}$
will be exactly the same as the descent associated with $L^{4p-1}_{CS}$, where 
each element $a^{q,n-q}$ is obtained from the corresponding one in the 
descent for $L^{4p-1}_{CS}$ upon taking the wedge product with 
$f_{K_{1}}\ldots f_{K_{m}}\,$. In other words, the products of the type $f_{K_{1}}\ldots f_{K_{m}}\,$ are completely spectators in a descent of $s_{W}$ modulo $d\,$.
That the $f_{K}$'s are $s_{W}$-closed is trivial once one realizes the identity
${\rm Tr}(R^{m(K)})\equiv {\rm Tr}(W^{m(K)})$ that is obtained from the relation 
$R^{ab} = W^{ab}+2e^{[a}P^{b]}$ 
where $e^{a}$ are the vielbein 1-forms and 
$P^{a}$ is the Schouten 1-form. 
\vspace{.3cm}

The lemmas 1, 2 and 3 give the general structure of the global conformal invariants 
in arbitrary dimensions, structure that we summarise in the following theorem.
\vspace{.3cm}

\noindent
{\underline{\it{Theorem\,:}}} ~
The global conformal invariants decompose into the integral, over the (pseudo)Riemannian 
manifold ${\cal M}_{n}(g)$, of linear combinations of strictly Weyl-invariant scalar densities 
and scalar densities that are invariants only up to a total derivative. 
The latter type of invariants further split into two distinct classes:
The integral of the Euler density in even dimension 
$n=2m$ and the integral of scalar densities of the type \eqref{Dragon}, 
that exist only in dimensions $n=4p-1\,$, $p\in \mathbb{N}^{*}\,$.

\section{Field equations for the Lorentz-Chern-Simons Lagrangians}
\label{sec:EOMLCS}

Given a pseudo-Riemannian spacetime ${\cal M}_{4p-1}$ of dimension $n=4p-1$ 
with an orientation, we consider the following functional of the metric, 
\begin{align}
I[g_{\mu\nu}] = \frac{1}{2p}\,\int_{{\cal M}_{4p-1}} L^{4p-1}_{CS}\;,
\end{align}
where $L^{4p-1}_{CS}$ is the Lorentz-Chern-Simons
$(4p-1)$-form descending from the following characteristic polynomial:
\begin{align}
{\rm Tr}\,R^{2p} = d L^{4p-1}_{CS}\;.
\end{align}
From the action functional $I[g_{\mu\nu}]$ we find, following the same steps as in \cite{Deser:1981wh} but generalised to higher
dimensions, the variational derivative
\begin{align}
{\cal E}^{\mu\nu}:= \frac{\delta I}{\delta g_{\mu\nu}} \equiv \frac{1}{2^{2p-1}}
\nabla^{\lambda}{\cal A}^{(\mu|\nu)}{}_{\lambda}\;,
\end{align}
where
\begin{align}
{\cal A}^{\mu|\nu}{}_{\lambda} := 
\varepsilon^{\mu \nu_{2}\nu_{3}\ldots\nu_{4p-1}}\;
[R_{\nu_{2}\nu_{3}}\ldots R_{\nu_{4p-2}\nu_{4p-1}}]^{\nu}{}_{\lambda}\;.
\end{align}
and $[R_{\nu_{2}\nu_{3}}\ldots R_{\nu_{4p-2}\nu_{4p-1}}]^{\nu}{}_{\lambda}$ 
denotes the $(2p-1)$-fold product of the 2-form valued matrix  
$[R_{\nu_{2}\nu_{3}}]^{\alpha}{}_{\beta} \equiv R^{\alpha}{}_{\beta\nu_{2}\nu_{3}}\,$.
Notice that the tensor density 
${\cal A}^{\mu|\nu\lambda}:= g^{\lambda\sigma}{\cal A}^{\mu|\nu}{}_{\sigma}$
is antisymmetric in the indices $\nu$ and $\lambda\,$.

Weyl invariance of the action $I[g_{\mu\nu}]$ gets translated into the following 
Noether identity
\begin{align}
g_{\mu\nu}{\cal E}^{\mu\nu}& \equiv\;0\;,
\end{align}
which is easily checked by using the algebraic Bianchi identity 
$R^{\alpha}{}_{[\mu\nu\rho]}{}\equiv 0\,$.
On the other hand, the fact that the Lorentz-Chern-Simons $(4p-1)$-form $L^{4p-1}_{CS}$ satisfies the equation $s_{D}L^{4p-1}_{CS}+db^{1,4p-2}=0$
is translated into the following Noether identity for the diffeomorphims:
\begin{align}
\nabla_{\mu}{\cal E}^{\mu\nu}& \equiv\;0\;.
\end{align}
This is less simple to show compared to the Noether identity for Weyl invariance,
as it rests on the identity
\begin{align}
\varepsilon^{\nu_{1}\ldots \nu_{4p-1} }\,{\rm Tr} [ R_{\nu_{1}\nu_{2}} \ldots R_{\nu_{4p-3}\nu_{4p-2}}R_{\nu_{4p-1}\nu}] \equiv 0\;,
\end{align}
which in its turn can be proved by using the obvious fact that the antisymmetrisation 
over $n+1$ indices in dimension $n$ gives identically zero and  the cyclicity 
of the trace over matrices. 

The last property that the tensor density ${\cal E}^{\mu\nu}$ satisfies is that, 
upon lowering 
one of its indices, it is strictly invariant under Weyl rescaling of the metric, or change 
of representative in the conformal class:
\begin{align}
s_{W} {\cal E}^{\mu\nu} = -2\,\omega \,{\cal E}^{\mu\nu} ~\Leftrightarrow~
 s_{W}{\cal E}^{\mu}{}_{\nu}   = 0\;.  
\end{align}
It is straightforward to show the above relations, for which it 
is useful to notice that one can rewrite the tensor density 
${\cal A}^{\mu|\nu}{}_{\lambda}$ as
\begin{align}
{\cal A}^{\mu|\nu}{}_{\lambda} := 
\varepsilon^{\mu \nu_{2}\nu_{3}\ldots\nu_{4p-1}}\;
[W_{\nu_{2}\nu_{3}}\ldots W_{\nu_{4p-2}\nu_{4p-1}}]^{\nu}{}_{\lambda}\;,
\end{align}
which is rather obvious, once one notices the identity   
\begin{align}
{\rm Tr}[R^{2m}] \equiv {\rm Tr}[W^{2m}]
\end{align}
relating the two $4m$-forms on both sides.
We recall that the matrix 
$[R]$ denotes the matrix-valued $2$-form $\frac{1}{2}\,dx^{\mu}dx^{\nu}[R_{\mu\nu}]\,$, with 
$[R_{\mu\nu}]^{\alpha}{}_{\beta} =R^{\alpha}{}_{\beta\mu\nu}\,$, and 
similarly for the matrix-valued Weyl 2-form
$[W]=\frac{1}{2}\,dx^{\mu}dx^{\nu}[W_{\mu\nu}]\,$.

\vspace{.3cm}

\section{Conclusions}
\label{sec:Ccl}

In this paper, we have given a general classification of global conformal invariants, 
completing previous works (see \cite{AlexakisBook} and references therein) 
where the conformal invariants related to the Lorentz 
Chern-Simons densities had been omitted. As a consequence of our decomposition 
of global conformal invariants, 
we see that the latter are \emph{not} in one-to-one
correspondence with the conformal anomalies, classified 
in \cite{Boulanger:2007ab,Boulanger:2007st}. Indeed, multiplying the Lorentz 
Chern-Simons densities by the Weyl factor does not produce any consistent conformal
(alias Weyl or trace) anomaly.  

We followed a purely cohomological method along the lines of the Stora-Zumino \cite{Sto84,Zumi84a,ManeStorZumi85a} treatment of anomalies in local Quantum Field Theory, where the BRST differential plays a central role.  
The descent equations in $H^{*,*}(s_{W}|d)$ starting with the $4p\,$-form
${\rm Tr}R^{2p}$ (in the universal algebra) for a pseudo-Riemannian 
manifold of dimension $4p-1$ 
gives a candidate for a consistent Weyl anomaly in a submanifold of codimension 1, 
i.e., an $(4p-1)$-form at ghost number 1 that is a nontrivial cocycle of the BRST 
differential $s_{W}$ modulo a $d$-exact term. 
Moreover, the variational derivative, with respect to the 
metric, of the Lorentz-Chern-Simons 
$(4p-1)$-form $L^{4p-1}_{CS}$ gives a symmetric rank-two tensor density 
${\cal E}^{\mu\nu}\equiv{\cal E}^{\nu\mu}$  with the  property that 
${\cal E}^{\mu}{}_{\nu}=g_{\nu\rho}{\cal E}^{\mu\rho}$
is not only divergenceless and traceless, but also exactly invariant 
under arbitrary rescalings of the metric.  

\section*{Acknowledgments}

N.B. is Senior Research Associate of the F.R.S.-FNRS (Belgium) and wants to 
thank G. Barnich, Th. Basile, F. Bastianelli and E. Joung for discussions.
He is also grateful to CPT for warm hospitality and to CNRS for financial support 
in the early stage of this work.
His work was supported in part by the F.R.S.-FNRS PRD grant ``Gravity and Extension'' 
number T.1025.14.
The project leading to this publication has also received funding from Excellence Initiative of 
Aix-Marseille University - A*MIDEX and Excellence Laboratory Archimedes LabEx, 
French ``Investissements d'Avenir'' programmes.

\providecommand{\href}[2]{#2}\begingroup\raggedright\endgroup

\end{document}